\documentclass[conference]{IEEEtran}
\IEEEoverridecommandlockouts
\usepackage{cite}
\usepackage{amsmath,amssymb,amsfonts}
\usepackage{algorithmic}
\usepackage{graphicx}
\usepackage{textcomp}
\usepackage{orcidlink}
\usepackage{booktabs}
\usepackage{acronym}
\usepackage{siunitx}
\usepackage{xcolor}
\usepackage{cleveref}
\usepackage{balance}
\usepackage{flushend}
\usepackage{dblfloatfix}    

\acrodef{AI}{artificial intelligence}
\acrodef{AoA}{angle of arrival}
\acrodef{CAD}{computer-aided design}
\acrodef{CIR}{channel impulse response}
\acrodef{CSI}{channel state information}
\acrodef{GAN}{generative adversarial networks}
\acrodef{GNSS}{global navigation satellite systems}
\acrodef{IPS}{indoor positioning systems}
\acrodef{ITS}[ITS]{intelligent transportation system}
\acrodef{mmWave}{millimeter-wave}
\acrodef{ML}{machine learning}
\acrodef{MPC}{multipath component}
\acrodef{LBS}{location-based services}
\acrodef{NLOS}{non line-of-sight}  
\acrodef{RMS}{root mean square}
\acrodef{RSSI}{received signal strength indicator}
\acrodef{RTLS}{real-time locating system}
\acrodef{TDoA}{time difference of arrival}
\acrodef{THz}{terahertz}
\acrodef{ToF}{time of flight}
\acrodef{SNR}{signal-to-noise ratio}
\acrodef{UWB}{ultra-wideband}
\acrodef{V2X}{vehicle-to-everything}
\acrodef{WSN}[WSN]{wireless sensor network}

\def\BibTeX{{\rm B\kern-.05em{\sc i\kern-.025em b}\kern-.08em
    T\kern-.1667em\lower.7ex\hbox{E}\kern-.125emX}}
\begin{document}

\title{Potentials of Deterministic Radio Propagation Simulation for AI-Enabled Localization and Sensing\\
}
\author{\IEEEauthorblockN{
Albrecht Michler\orcidlink{0000-0002-3434-3488},
Jonas Ninnemann\orcidlink{0000-0001-7988-079X},
Jakob Krauthäuser, 
Paul Schwarzbach\orcidlink{0000-0002-1091-782X},
Oliver Michler\orcidlink{0000-0002-8599-5304}}

\IEEEauthorblockA{Institute of Traffic Telematics, Faculty of Transport and Traffic Sciences "Friedrich List" \\
Technische Universit\"at Dresden\\
Email: \{firstname.lastname\}@tu-dresden.de}
}

\maketitle

\begin{abstract}
Machine leaning (ML) and artificial intelligence (AI) enable new methods for localization and sensing in next-generation networks to fulfill a wide range of use cases. These approaches rely on learning approaches that require large amounts of training and validation data. 
This paper addresses the data generation bottleneck to develop and validate such methods by proposing an integrated toolchain based on deterministic channel modeling and radio propagation simulation. The toolchain is demonstrated exemplary for scenario classification to obtain localization-related channel parameters within an aircraft cabin environment. 
\end{abstract}


\section{Introduction}


\Ac{LBS} significantly increase the potential and application scenarios of wireless systems by extracting geometric information from radio signals to determine the location of a user or object. \Ac{LBS} are powered by \ac{GNSS} or \ac{IPS} \cite{zafariSurveyIndoorLocalization2019, mendozasilvaMetaReviewIndoorPositioning2019} to provide a solution for various tasks, such as localization, tracking, counting of objects and people. In addition to an device-based active localization, the integrated use of \acp{WSN} also enables device-free radio sensing functionalities for radar-like imaging and object detection \cite{shastriReviewMillimeterWave2022a}. 
Concurrently, techniques for \ac{ML}, as a branch of \ac{AI}, have become crucial in next-generation wireless communication systems such as 6G\cite{wang6GWirelessChannel2020}. \ac{ML} techniques can improve conventional methods relating to channel modeling \cite{huangArtificialIntelligenceEnabled2022}, channel measurements \cite{wang6GWirelessChannel2020}, antenna-channel optimization \cite{huangArtificialIntelligenceEnabled2022a}, wireless networking \cite{challitaWhenMachineLearning2020} and fingerprinting \cite{tsengRayTracingAssistedFingerprintingBased2017}. But as stated in \cite{huangArtificialIntelligenceEnabled2022a} one key challenge remains: the generation of massive amount of data for learning and training. This necessitates specific hardware and expensive, time-consuming measurement campaigns.

\begin{figure}[t]
    \centering
    \includegraphics[width=0.9\linewidth]{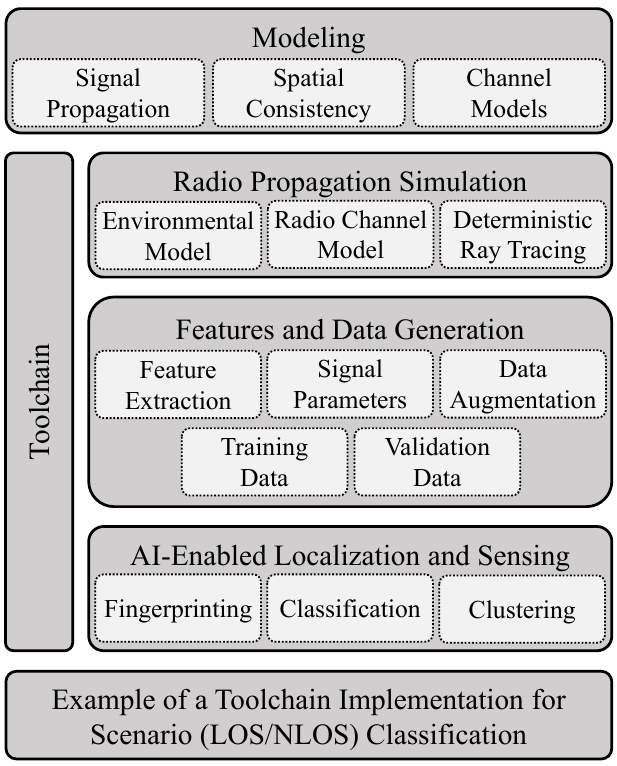}
    \caption{Structure of the paper to enable data generation by using deterministic radio propagation simulation for \ac{AI}-enabled localization and sensing.}
    \label{fig:structure}
\end{figure}

This paper addresses the data generation bottleneck for AI-enabled localization by proposing a toolchain based on deterministic channel modeling and radio propagation simulation to generate \ac{CSI}. This approach is particularly useful for high-mobility scenarios and specific environments, as it facilitates the development and initial validation of AI-enabled localization and sensing methods. These methods can be applied in various applications, such as \ac{ITS}, logistics, localization, object detection and counting. The overall structure of the paper is depicted in \cref{fig:structure}.

\section{Problem outline}
\subsection{Data generation for AI-based localization}

AI-based approaches for localization and sensing employ machine-learning and artificial intelligence methods to improve accuracy, efficiency and integrity in determining the location of devices or sensing the environment. Localization algorithms utilize various approaches that can operate on raw channel state information data (e.g. \ac{CIR}), derived data (e.g. range) or during the sensor fusion stage (e.g. integrating radio-based localization and inertial navigation). AI-based channel models offer a key advantage over conventional models by exhibiting high adaptability to different environments, enhancing the overall robustness of the localization process. However, in order to apply such models within the radio localization domain, a learning procedure must be performed first. This generally involves the following steps: 
\begin{enumerate}
    \item Data collection or generation
    \item Data annotation
    \item Training
    \item Validation
    \item Deployment
\end{enumerate}

Within this sequence, data generation or collection plays a critical role. Existing literature indicates that low-quality training data limits the overall performance of the system \cite{huangArtificialIntelligenceEnabled2022a}. Multiple approaches can be employed for data collection or generation. The most straightforward is to gather sensor data within the designated deployment area. This method captures the full complexity of physical effects within the data. On the contrary, live measurement campaigns can be time consuming and the data is scenario-specific and therefore may not generalize well. Furthermore, calculating associated ground truth values can be challenging, making it harder to label the data correctly.

To address these challenges, simulation methods can be utilized. Such methods provide synthetic data that closely resemble real-world data while offering certain advantages. The primary advantage lies in the ability to generate an arbitrary number of data samples in a short time, making this approach more efficient than measurement-based data generation. Furthermore, data can be labeled and assigned ground-truth values directly.

\subsection{Sensor specifics for radio based localization}
There is a variety of sensors, that are generally used for localization and sensing. These include radio based systems, optical camera systems, laser-based systems and inertial measurement systems. When using one or multiple sensor systems in a single-sensor or data-fusion localization system, it is essential to generate matching training data to establish a consistent AI-based localization system.

For optical systems, \ac{GAN} and data augmentation on existing data sets can be applied to generate new training data \cite{gan}. Inertial measurements can be generated by calculating motion forces along a predefined trajectory and adding sensor noise according to a sensor specification \cite{imusim}. Laser-based or quasi-optical radio localization, which relies on distance or angle estimation, can be simulated using simple ray-tracing approaches, where \ac{NLOS} paths can be omitted due to shading. 

In contrast, for radio-based localization, a variety of primary measures based on \ac{CSI} can be used. This include \ac{ToF}, \ac{RSSI}, \ac{SNR} or \ac{AoA} estimates. Especially in high-dynamic scenarios, models need to be able to generate spatially consistent signal parameters, since real channel information exhibits sensitivity to spatial and temporal correlations \cite{8647188}. As a result, solely applying stochastic channel models may lead to unrealistic and overly simplistic data, which in turn leads to mis-trained AI-based models. The impact of spatial consistency is further discussed in \cref{ssec:spatial}.  Additionally, running real-world measurement campaigns poses challenges as it requires the availability of hardware samples. For newly specified radio technologies, off-the-shelf options are often unavailable and need to be specifically designed and manufactured.

Deterministic radio propagation simulation addresses these challenges by providing the necessary channel and signal parameters for a desired environment, radio properties and antenna parameters.

\section{Fundamentals and Modeling}

\subsection{Signal Propagation}
\label{ssec:signal_prop}

The radio channel consists of the transmitting antenna, the propagation channel and the receiving antenna. While propagation, the electromagnetic wave encounters different objects in the environment causing the three basic propagation phenomena: reflection, diffraction, and scattering. This leads to various channel characteristics, which can be categorized into large-scale and small-scale fading effects. Large-scale fading is caused by the change in signal strength over distance due to path loss and shadowing by obstacles. Small-scale fading, on the other hand, is caused by multipath propagation and constructive or destructive interaction and interference of the electromagnetic wave during propagation.

In the case of multipath propagation, the signal reaches the receiver via multiple paths. When the direct line-of-sight (LOS) path between transmitter and receiver is obstructed, non-line-of-sight (NLOS) propagation can occur. In this case, the received signal is composited of different delayed, attenuated, and phase-shifted waves from various directions. This results in various \acp{MPC} in the \ac{CIR}. This multipath fading and \ac{NLOS} propagation have severe effects on the localization performance, because the \ac{ToF} and range are determined incorrectly. To address this source of error various \ac{ML} algorithms for scenario classification of LOS/\ac{NLOS} is applied (cf. \cref{sec:example}).

\subsection{Spatial Consistency}
\label{ssec:spatial}
Spatial consistency enables channel models to provide spatially consistent and time-evolving \acp{CIR} for different sensor locations and environments. Therefore, spatial consistency is crucial for evaluation of localization and sensing in a specific use cases. However, most current statistical channel models are drop-based, which are only able to generate \acp{CIR} for a particular user at a randomly chosen location and provide no spatial correlation between consecutive simulation runs \cite{rappaportWirelessCommunicationsApplications2019}. This is a limitation for AI-based localization approaches, which rely of full CSI in order to infer geometric relations of the scenario. Therefore, the goal is to generate smoothly time-evolving \acp{CIR} based on the user movement in high-mobility scenarios, such as \ac{V2X} communications. This way, AI-enabled localization and sensing methods can be trained and validated using the desired user motion and environment. Additionally, in \ac{mmWave} and \ac{THz} band the narrow antenna beams results in highly correlated channel characteristics \cite{shastriReviewMillimeterWave2022a}.

\subsection{Channel Models}
In principle, three different types of channel models can be distinguished: deterministic, stochastic, and hybrid models. \Cref{tab:channel_models} compares these type of channel models in terms of their properties, requirements, and available simulators.

\begin{table}[ht]
\caption{Comparison of types of channel models.}
\begin{center}
\begin{tabular}{p{1.6cm} p{1.9cm} p{1.8cm} p{1.8cm}}
        \toprule
		\textbf{}   & \textbf{Deterministic}   & \textbf{Stochastic} & \textbf{Hybrid}\\ 
		\midrule
		Requirements and Inputs    & Geometry of the environment, electromagnetic properties of the materials and spatial position of the sensors & statistical approximations using random distributions of the channel parameters & geometry of the propagation environment and random distributions of the channel parameters \\\midrule
		Spatial consistency    & Yes & No  &  Yes, in some cases  \\\midrule
		Complexity & High  & Moderate & Medium\\\midrule
		Accuracy  & High  & Moderate & High (static) \\\midrule
		Available Simulators  & NYURay, CloudRT, Altair WinProp & NYUSIM  & QuaDRiGa \\	\bottomrule
\end{tabular}
\vspace{-0.3cm}
\label{tab:channel_models}
\end{center}
\end{table}

\textbf{Deterministic channel models} solve Maxwell's equations numerically in a given geometric environment. They describe the channel and temporal variations due to the number, position, and characteristics of reflectors in the environment.  One example of a deterministic algorithm is Ray Tracing (RT), where the individual propagation paths or rays are calculated individually based on the channel characteristics. \cite{wang6GWirelessChannel2020, hanTerahertzWirelessChannels2022}

\textbf{Stochastic channel models} use statistical approximations, employing random distributions to characterize the received signal and channel parameters such as path loss, delay, number of paths, and fading. These models are primarily based on measurements and empirical observations in specific types of environments (rural, urban, indoor, micro, macro, etc.), rather than on the position of the sensors.   Therefore, the channel is described at a random location of the sensor in a defined type of environment, leading to a lack of spatial consistency between multiple simulation runs and different sensor locations \cite{wang6GWirelessChannel2020, hanTerahertzWirelessChannels2022, rappaportWirelessCommunicationsApplications2019}.

\textbf{Hybrid channel models} combine deterministic and stochastic approaches and therefore offer a balance between accuracy and complexity. Geometry-based stochastic model (GBSM), such as the 3GPP TR 38.901 \cite{3gpp2019technical} model and the WINNER II model \cite{kyosti20074}, incorporate the geometry between transmitter and receiver through the length and angle of the LOS signal component. However, they still use stochastic models and distributions to describe the various signal parameters. Quasi-deterministic channel models compute the dominant propagation path with a highly simplified environment map and add clusters of stochastic modeled MPC to the model. Such models are only partially spatial consistency and therefore not suitable for the evaluation localization and sensing functionalities in a specific environment and use case. \cite{hanTerahertzWirelessChannels2022}

\section{Toolchain}
\label{sec:toolchain}

\begin{figure}[ht]
    \centering
    \includegraphics[width=1\linewidth]{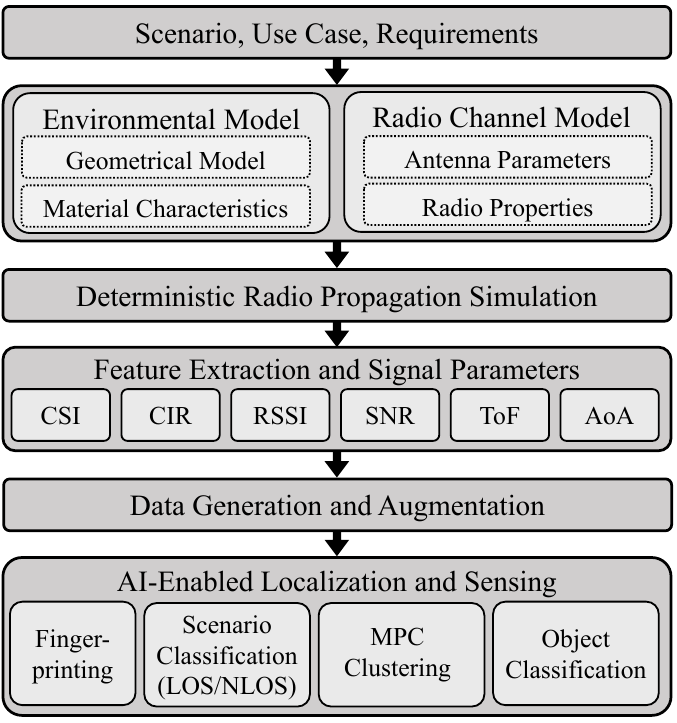}
    \caption{Toolchain for data generation with deterministic radio propagation simulation to support \ac{AI}-enabled localization and sensing. }
    \label{fig:toolchain}
\end{figure}

The overall toolchain describes the process from scenario definition and modeling, to running the radio propagation simulation, extracting relevant parameters, generating signal data and finally training and evaluating the AI model. A schema of the toolchain is depicted in \cref{fig:toolchain}. 

\subsection{Scenario Modeling}
In order to generate data for AI applications, it is necessary to define the use case and scenario. To facilitate data generation using deterministic simulation, it is important to model the properties of the radio channel and the environment in as much detail as possible. This involves two main aspects: Firstly, the geometry of the physical surroundings needs to be accurately represented within the simulation framework, for example in form of a 3D \ac{CAD} model. This entails, that electromagnetic material constants (permittivity, permeability) for all elements are provided in order to properly simulate all propagation effects. Secondly, the radio channel and the simulated hardware need to be parameterized. This includes the center frequency, transmit power, antenna gain, 3D antenna pattern, and other simulation parameters. 

\subsection{Deterministic Radio Propagation Simulation}
The environmental model and radio channel model are used for the 3D deterministic radio propagation simulation to compute all propagation paths or rays of the radio wave, between the transmitter and the desired receiver location. This deterministic approach takes the effect of the environment on the propagation into account accurately. In addition, multiple parameters, such as path loss, \ac{RSSI} \ac{CIR}, \ac{AoA}, and \ac{ToF} could be predicted with one simulation run. 

The radio rays are computed considering refraction, reflection, diffraction or scattering either by ray tracing and ray launching. Ray tracing determines the individual paths backwards from receiver to the transmitter and ray launching launches a number of rays from the transmitter and calculate their paths from there. With a time-variant simulation it is also possible to compute the signal propagation in a dynamic scenario based on a trajectory.

Various simulators exist for 3D ray tracing and ray launching, such as NYURay for \ac{mmWave} \cite{rappaportWirelessCommunicationsApplications2019}, CloudRT \cite{heDesignApplicationsHighPerformance2019}, MaxRay \cite{arnoldMaxRayRaytracingbasedIntegrated2022a}. In this case, the standard ray tracing model of Altair WinProp 2022.2.2 is employed to simulate signal propagation and generate corresponding signal parameters \cite{AltairWinProp2022}. 

\subsection{Feature Extraction and Signal Parameter}
A variety of signal parameters can be leveraged for localization and sensing purposes, ranging from \ac{CSI} to secondary parameters. Channel estimation is the method used to obtain the \ac{CSI} from a wireless communication link.  CSI represents the properties and parameters of the channel, describing how the signal propagates from the transmitter to the receiver. The CIR can indicate the instantaneous channel conditions and consists of Multipath Components (MPCs) resulting from the set of all propagation paths.

Up until this point, all simulation steps are deterministic and can directly be reconstructed. However, one simulation step yields only one data sample, which does not address the issue of a high number of samples needed for AI training. Furthermore, some physical parameters are presented in an idealized form. Therefore, it is necessary to reconstruct the physical channel properties.

\subsection{Data Generation and Augmentation}
The \ac{CIR} $h(\tau)$ is obtained from the ray tracing simulation consisting of multiple Dirac pulses $\delta$ at a certain propagation delay $\tau_i$ and amplitude $a_i$ of the $i$-th propagation path or \ac{MPC} computed from the length of the ray path and the path loss. Mathematically, the \ac{CIR} $h(\tau)$ of the time-invariant propagation channel can be described as following:
\begin{equation}
	h(\tau) = \sum_{i=1}^{N} a_i \delta(\tau-\tau_i)
	\label{equ:cir_time_invariant}
\end{equation}

\begin{figure}[ht]
    \centering
    \includegraphics[width=1\linewidth]{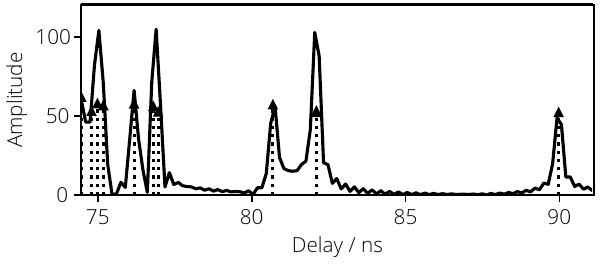}
    \caption{Band-limited reconstructed \ac{CIR}.}
    \label{fig:cir_bandlimited}
\end{figure}

The CIR obtained from the simulation assumes an unlimited bandwidth. However, in real wireless systems, the channel is band-limited by a bandwidth $B$. Therefore, the \ac{CIR} needs to be reconstructed with band-limited conditions by applying the Whittaker–Shannon interpolation formula or sinc interpolation. This results in \acp{MPC} with a certain width and a limited range resolution, which are important for localization and sensing purposes \cite{schwarzbachEnablingRadioSensing2022}. An exemplary reconstruction of a \ac{CIR} obtained in \ac{NLOS} conditions is shown \cref{fig:cir_bandlimited}.

More signal parameters and features for \ac{ML} methods can be derived from the \ac{CIR} and the deterministic simulation. These features include \ac{RSSI}, \ac{SNR}, \ac{AoA}, \ac{ToF}, \ac{TDoA}, mean excess delay, \ac{RMS} delay spread, and kurtosis.

To generate multiple data samples from a single simulation step, the interpolation filter properties, such as bandwidth, can be varied. Furthermore, convolution with a skewed noise function is possible to introduce additional variation. These techniques help prevent overfitting to a specific scenario.

\begin{figure*}[!b]
    \centering
    \vspace{-0.3cm}
    \includegraphics[width=.95\textwidth]{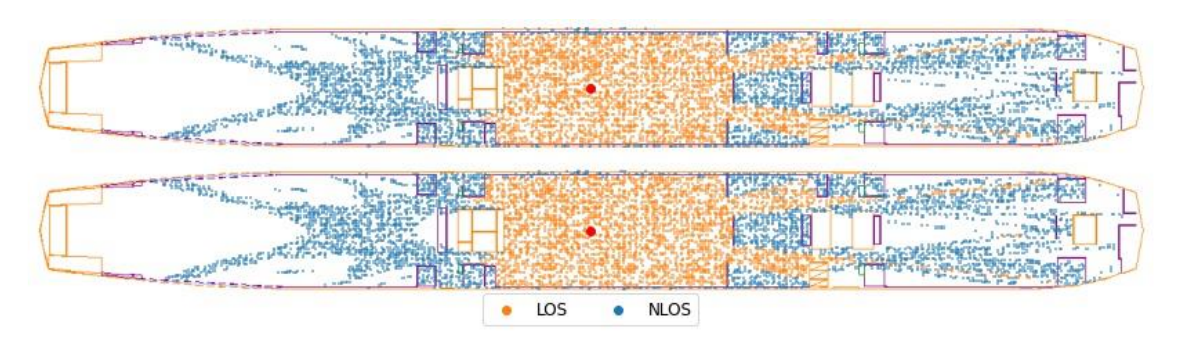}
    \caption{Scenario classification (LOS/\ac{NLOS}) in the aircraft cabin. Ground truth obtained from the radio propagation simulation with ray tracing (top) and estimate by the random forest classifier (bottom). Sender position is marked with a red dot.}
    \label{fig:classification_aircraft}
    \vspace{-0.8cm}
\end{figure*}
 
\subsection{AI-Enabled Localization and Sensing}

Data driven methods enable a variety of applications in modern localization and sensing models. They can be applied at different levels of signal properties and features. \Cref{tab:methods} lists the several \ac{ML} task and methods related to localization and sensing.

\begin{table}[ht]
\caption{Overview of \ac{ML} Methods for localization and sensing.}
\label{tab:methods}
\begin{tabular}{p{3cm} p{2.8cm} p{1.5cm}}
\toprule
\textbf{\ac{ML} Task and Method}                 & \textbf{Features}                                        & \textbf{Reference} \\ \midrule
\ac{MPC} Detection and Clustering       & \ac{CIR}                                                      &   \cite{wang6GWirelessChannel2020}        \\ \midrule
Fingerprinting                     & \ac{RSSI}, \ac{CIR}                                                &  \cite{wang_novel_2020} \cite{gong_transformer-based_2023}       \\ \midrule
Scenario (LOS/\ac{NLOS}) Classification & \ac{CIR}, \ac{RSSI}, \ac{RMS} delay spread, kurtosis, mean excess delay &  \cite{Feature-based} \cite{huangArtificialIntelligenceEnabled2022}   \cite{maranoNLOSIdentificationMitigation2010}     \\ \midrule
Outlier Detection                  & Range, \ac{AoA}                                               &     \cite{chukuRSSIBasedLocalizationSchemes2021}      \\ \midrule
Occupancy Detection and Object Classification               & \ac{CIR}, \ac{MPC}                        &    \cite{ninnemannMultipathassistedRadioSensing2022} \\
\bottomrule
\end{tabular}
\end{table}

For each \ac{ML} task (classification, clustering, detection) there are several algorithms available to solve the corresponding problem. In general, this algorithms can be divided into supervised, unsupervised and reinforcement learning methods.

\section{Example for Scenario (LOS/NLOS) Classification}
\label{sec:example}

The toolchain described in \cref{sec:toolchain} is exemplarily applied to generate data for a scenario (LOS/\ac{NLOS}) classification using supervised learning. For the purpose of active localization and evaluating scenario classification, a use case in the area of Intelligent Transportation Systems (ITS) is chosen. Specifically, the toolchain is evaluated in the context of the connected aircraft cabin, considering the significant potential of \ac{LBS} in this environment. Examples of potential applications include passenger boarding/deboarding, technology-based social distancing methods ({COVID}-19) \cite{schwarzbachEvaluationTechnologySupportedDistance2020b} and object detection. Localization methods based on \ac{RSSI} or \ac{ToF} for distance estimation are heavily influenced by the environment due to the signal multipath reception in the aircraft cabin, where scenario classification yields potential for improving localization accuracy and robustness. Additionally, the limited accessibility of aircraft cabins can hinder extensive real-world measurement campaigns. 

For scenario, a 3D \ac{CAD} model of the Airbus A340 cabin was utilized. The study is aimed at identifying areas with LOS/NLOS coverage given a fixed anchor position. As radio access technology, \ac{UWB} was chosen, with a center frequency of (\SI{3500}{\mega \hertz}) and maximum transmit power of –16 dbm. \ac{UWB} as radio technology for localization and tracking is widely used and achieves a very high ranging accuracy (\SI{10}{\centi \meter}) due to the high bandwidth (\SI{500}{\mega \hertz}). This \ac{RTLS} is suitable for an aircraft cabin use-case \cite{karadenizPreciseUWBBasedLocalization2020}. Both transmission and reception utilized omnidirectional antennas in the simulation.

\begin{figure}[t]
    \centering
    \includegraphics[width=1\linewidth]{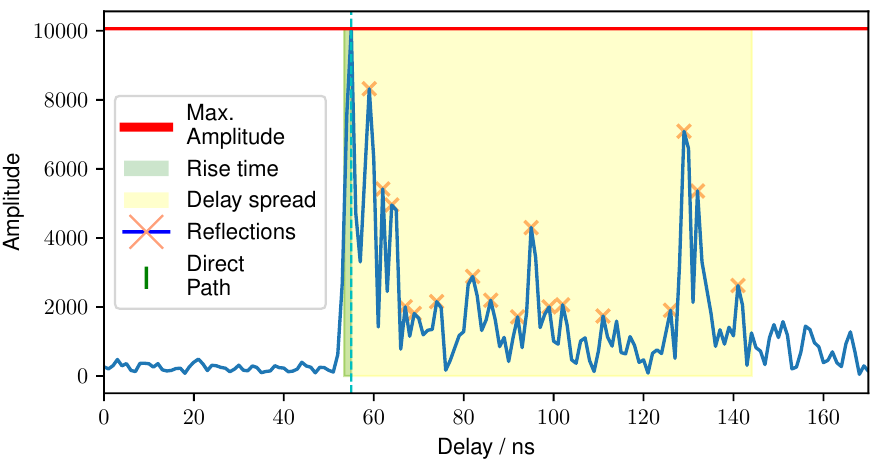}
    \caption{Parameters obtained from a channel impulse response (CIR) sample.}
    \label{fig:cir_parameter}
\end{figure}

For the scenario classification, various features are extracted from the reconstructed \ac{CIR}. These features included \ac{RMS} delay spread, amplitude, kurtosis, total received energy, mean excess delay, maximal amplitude, and \ac{RSSI} (see \cref{fig:cir_parameter}).  Recursive feature elimination was employed to select the most relevant and suitable features in the training dataset.
\balance

Data in form of the \ac{CIR} was generated for one transmit antenna position in the middle of the aircraft and a grid of possible receiver locations. The random forest classifier was used as the machine learning algorithm for scenario classification.

The results of the classification were presented and compared to the ground truth obtained from the radio propagation simulation with ray tracing (see \cref{fig:classification_aircraft}). The overall classification accuracy achieved using the random forest model was \SI{98.51}{\percent}.

\section{Conclusion}
The paper presents a developed toolchain for generating data for AI-driven localization and sensing. Compared to real-world data generation, the toolchain offers advantages in terms of accessibility, reproducibility, and the availability of ground truth for verification. The main advantage compared to stochastic channel modelling lies in the preservation of spatial consistency, which in turn leads to a more realistic and accurate channel simulation and enables the utilization of the full channel state information in localization and sensing algorithms. The toolchain's effectiveness is demonstrated through a UWB LOS/NLOS classification study conducted in an aircraft cabin. This approach can be extended and generalized to other environments and radio access technologies. In order to verify models, real-world tests can be conducted. This is particularly important to identify any biases that may arise during the simulation-based learning phase. Furthermore, a correction framework can be developed to address such biases or unknown effects observed during the real-world validation process and feed back corrected parameters to the original model.

There is potential for automating scenario design using machine learning techniques, although careful consideration must be given to bias propagation within such a setup. Additionally, the application layer of the toolchain can be expanded and validated with other machine learning/AI methods, such as deep learning methods.
\newpage
\section*{Acknowledgment}

\begin{tabular}{p{6.cm} p{3.6cm}}
This work has been funded by the German Federal Ministry for Economic Affairs and Climate Action (BMWK) following a resolution of the German Federal Parliament within the projects CANARIA (FKZ: 20D1931C) and INTACT (FKZ: 20D2128D). & 
\hspace{-0.5 cm} \raisebox{-0.8\height}{\includegraphics[width=0.7\linewidth]{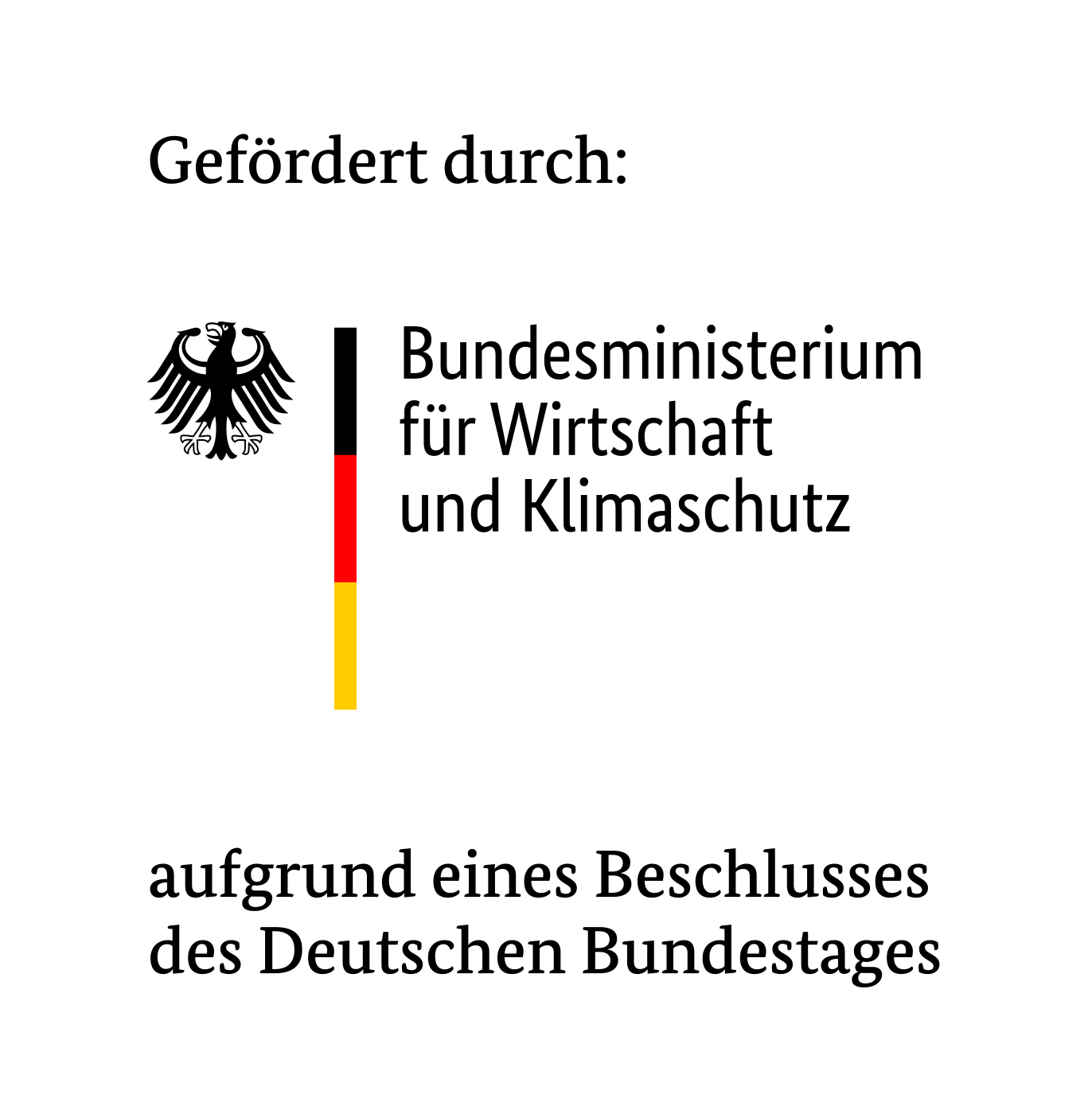}} \\
\end{tabular}

\bibliographystyle{IEEEtran}
\bibliography{lit.bib}

\end{document}